\documentclass[reqno]{amsart}

\usepackage{amsmath}
\usepackage{amssymb}
\usepackage{amsthm}

\newcommand\C{{\mathbb C}}                

\newcommand\R{\mathbb R}

\newcommand\sA{{\stoch A}}

\newcommand\sB{{\stoch B}}

\newcommand\E{{\mathbb E}}

\newcommand\M{{\mathcal M}}          

\newcommand\sM{{\stoch M}} 
       
\renewcommand\L{{\mathcal L}}
\newcommand\cL{{\mathcal L}}

\newcommand\bP{{\mathbb P}}

\newcommand\Q{{\mathcal Q}}
\newcommand\cQ{{\mathcal Q}}

\renewcommand\t{{\mathcal T}}

\newcommand\meas{{m}}   

\newcommand\stoch{\mathsf}

\newcommand\Kato{{\mathcal K}}
\newcommand\Katopm{{\mathcal K_\pm}}
\newcommand\Kloc{{\mathcal K}_{\mbox{\scriptsize\it loc}}}

\newcommand\Bg{{L^2_{\mbox{\scriptsize\it hol}}(h\meas)}}
\newcommand\hol{{\mbox{\scriptsize\it hol}}}
\newcommand\loc{{\mbox{\scriptsize\it loc}}}
\newcommand\ol{\overline}
\renewcommand\Re{\mathfrak{Re}\,}
\renewcommand\Im{\mathfrak{Im}\,}
\newcommand\Cinfty{\mathop{\hbox{$C$}}^\infty}
\renewcommand\centerdot{\mathbf{\cdot}}

\def\downto{{\mathchoice
{\raise.25ex\hbox{$\,\scriptstyle\searrow\;$}} 
{\raise.25ex\hbox{$\,\scriptstyle\searrow\;$}} 
{\raise.25ex\hbox{$\scriptscriptstyle\searrow$}} 
{\raise.25ex\hbox{$\scriptscriptstyle\searrow$}} 
}}

\newcommand\abs[1]{\left\vert {#1}\right\vert}
\newcommand\ep[1]{{e}^{\textstyle #1}}
\newcommand\half{{\frac 1 2}}

\newcommand\Cov{\mathrm{Cov}}

\newcommand\up[1]{^{(#1)}}

\newcommand\eq[1]{(\ref{eq:#1})}

\theoremstyle{definition}
\newtheorem{definition}{Definition}[section]

\newtheorem{remarks}[definition]{Remarks}

\newtheorem{examples}[definition]{Examples}

\newtheorem{lemma}[definition]{Lemma}
\newtheorem{proposition}[definition]{Proposition}
\newtheorem{theorem}[definition]{Theorem}

\theoremstyle{remark}

\newtheorem*{itremark}{Remark}

\begin{document}

\title[Relating resolvents of Berezin-Toeplitz
operators]{A transformation formula relating resolvents of Berezin-Toeplitz
operators by an invariance property of Brownian motion}
\author{Bernhard G. Bodmann}
\date{\today}
\address{337 Jadwin Hall, Department of Physics, Princeton University, 
Princeton, NJ 08544, USA}
\email{bgb@princeton.edu}

\begin{abstract}
Using a stochastic representation provided by Wiener-regularized path integrals for the
semigroups generated by certain Berezin-Toeplitz operators, a transformation formula 
for their resolvents is derived. The key property used in the transformation of the 
stochastic representation is that, up to a time change, Brownian motion is invariant 
under harmonic morphisms. This result for Berezin-Toeplitz operators
is obtained in analogy with a well-known technique generating relations among 
Schr\"odinger operators that was recently generalized to Riemannian manifolds 
[Wittich, J.\ Math.\ Phys.\ {\bf 41} (2000), 244].  
\end{abstract}


\maketitle

\section{Introduction}
\noindent
The idea for the results presented here is taken from a transformation
formula relating resolvents of certain Schr\"o\-din\-ger operators \cite{DK79,Bla82,CS90}.
Recently, Wittich \cite{Wit00} proved this formula and a generalization in the setting 
of Riemannian manifolds with the help of an invariance property of Brownian motion 
under harmonic morphisms. 

An analogous strategy applied to the probabilistic representation of 
Berezin-Toeplitz semigroups by Wiener-regularized path integrals \cite[Sec.~4]{Bod}
gives a relationship between resolvents of different Berezin-Toeplitz 
operators whenever the base manifolds of two holomorphic line bundles 
have K\"ahler structures that are locally conformally equivalent.
Such relationships can link operators that model quantum mechanical 
systems with qualitatively different degrees of freedom, for example those
described by various Lie groups \cite{Per86,BMM96}.

An intermediate step in the derivation of the main result is to relate 
resolvents of Schr\"odinger operators arising as
perturbations of the Bochner Laplacian in complex line bundles. 
The technique of a stochastic time change in path-integral 
representations of such operators is already known \cite{Sto94}. 
Here, we combine harmonic morphisms with an appropriate time change
to exploit the invariance property of Brownian motion and to establish a relationship 
between resolvents of Schr\"odinger operators. The main result is then derived from 
the intermediate step by a monotone limit of 
certain Schr\"odinger operators.

Unfortunately, the rigidity of harmonic morphisms only allows 
a rather trivial result when the base manifolds have complex dimension 
higher than one \cite{Fug78}. Therefore, we restrict the discussion
to Riemann surfaces.

This paper is organized as follows: In Section~2, we fix the
notation and briefly explain the relevant background information. Section~3
contains the statement of the transformation formula as the main result and a few illustrating 
examples, followed by the proof.

\section{Basic Definitions and Concepts\label{sec:2}} 
\subsection{Hilbert spaces of square-integrable, holomorphic sections}%

\begin{definition} 
Let $\M$ be a Riemann surface, that is, a complex manifold
of dimension one, and let $\L$ be a holomorphic line bundle
over $\M$, equipped with a Hermitian metric $h=\{h_x\}_{x \in \M}$ 
on its fibers.
To be precise, for each base point $x \in \M$ there is a
sesquilinear metric $h_x: \L_x\times\L_x \to \mathbb C$ on the associated 
fiber $\L_x$ and by convention, each $h_x$ is conjugate linear in the first 
argument. Given a measure $\meas$ on $\M$,  
we may define an inner product
\begin{equation}
  \label{eq:IP}
  (\psi,\phi) := \int_\M h_x(\psi(x),\phi(x))\, d\meas (x)
\end{equation}
for sufficiently regular sections $\psi$ and $\phi$.

The linear space of sections in $\L$ will be denoted as $\Gamma_\L(\M)$.
The subspace of square-integrable sections on a complex line bundle 
$\L$ over a base manifold $\M$ is denoted by
\begin{equation}
  \label{eq:L2space}
  L^2(h \meas) := \Bigl\{ \psi \in \Gamma_\cL(\M): 
                         \int_\M h_x(\psi(x),\psi(x))\, d\meas(x) < \infty \Bigr\} \, .
\end{equation}
When $\L$ is a holomorphic line bundle, we define
the generalized Bergman space $L^2_{hol}(h \meas)$ as
the space of all holomorphic sections in $L^2(h \meas)$.
\end{definition}

\begin{remarks}
Equipped with the previously defined inner product, 
the space $L^2(h \meas)$ containing all square-integrable sections 
becomes a Hilbert space in the usual way by identifying
sections that do not differ up to sets of measure zero.

If $\L$ is a holomorphic line bundle and 
$\meas$, interpreted as a volume form, as well as
$h$ are everywhere non-degenerate 
and smooth, then the generalized Bergman space 
$\Bg$ is a space of functions that
may be identified with a Hilbert-subspace
of $L^2(h \meas)$ \cite[Sec.~2]{Bod}.
\end{remarks}

\begin{examples} \label{ex:I}
Unless otherwise noted, the generalized Bergman spaces
cited as examples may be obtained from unitary irreducible 
Lie group representations as described by Onofri \cite{Ono75}. 
\begin{enumerate}
\item \label{ex:IIa}
 \textit{Fock-Bargmann space}. 
        Let $\M=\C$ and $\L=\M \times \C$.
        The volume measure on $\M$ is simply the
        normalized Lebesgue measure
        $dm = d^2z/\pi = dz_1 dz_2/\pi$,
        where $z_1, z_2 \in \mathbb R$ denote the 
        real and imaginary parts of $z=z_1+i z_2$.
        Every vector at $z \in \M$ can be thought
        of as a pair $(z,u)$, $u \in \mathbb C$.
        The Hermitian metric on the fibers of $\L$ is defined 
        over a base point $z \in \C$ 
        by $h_z((z,u),(z,v))=e^{-\abs{z}^2} \ol u v$.
        The space $\Bg$ obtained in this setting is
        infinite-di\-men\-sion\-al, since the sections
        $z \mapsto (z,z^n)$, $n \in \mathbb N$ are
        square-integrable and pairwise orthogonal.
        This space is related to a representation of
        the Heisenberg-Weyl group \cite{Bar61}.
  \item \textit{Barut-Girardello space}. \label{ex:II}
        The preceding example with instead the
        Hermitian metric
        $h_z((z,u),(z,v))=e^{-\abs{z}} \ol u v$
        produces a space that is also infinite-di\-men\-sion\-al
        and relates to $SU(1,1)$ \cite{BG71}.
  \item \textit{Generalized Bergman space over 
                powers of the tautological bundle}. \label{ex:IIb}
   Let $\L^\times= \C^{2}\setminus\{0\}$ 
   and $\M = \C \mathrm{P}^1$ that is obtained by identifying two 
     nonzero vectors $w=(w\up{1},w\up{2})$ and $w'=({w'}\up{1} ,{w'}\up{2})$ 
     whenever they are collinear, $w = c w'$ for some $c \in \C\setminus \{0\}$.
    The equivalence class of $w$ will be written as $[w]$. We
   choose local coordinates given on the set
   $U_{1}:=\{[w]: w \in \C^{2}, w\up{1} \neq 0\}$ by
   $\phi_{1}([w]):=w\up{2}/w\up{1}$ and analogously 
   on $U_2$ by flipping $w\up 1$ and $w\up 2$.
   Filling in the missing zero in each fiber $\L_{[w]}$, that is,
   the vectors belonging to the equivalence class $[w]$, 
   would yield the so-called tautological line bundle. We consider,
   more generally, powers of this bundle by
   picking an integer $k \in \mathbb Z$ and 
   specifying the transition functions between the local trivializations
   over $U_1$ and $U_2$ as
   $t_{1,2}:=(w\up{2}/w\up{1})^k$ and $t_{2,1}$ with $w\up{1}$, $w\up{2}$
   exchanged. When $k < 0$, there is no global
   holomorphic section, which makes a nontrivial $\Bg$ impossible.
   However, if $k$ is zero or a positive integer, 
   the vector space of holomorphic sections is $k+1$-dimensional \cite{Wel80}.
   One may turn it into a Hilbert space by using
   the Hermitian metric from the inner product on $\C^2$ and choose
   the measure $m$ to be invariant under the action of complex automorphisms 
   on the base manifold. The action of $SU(2)$ on $\C^2$ yields for different 
   choices of $k \in {\mathbb Z}^+$ the unitarily inequivalent irreducible 
   $SU(2)$ representations on the corresponding generalized Bergman 
   spaces \cite{Per86}.
\item \textit{A space of theta functions as generalized Bergman space}. \label{ex:theta}
     Given a lattice ${\mathbb{G}}:=\{ l_1 \epsilon_1 + l_2 \epsilon_2: l \in {\mathbb{Z}}^{2}\}$
     with spacings $\epsilon_{1,2} \in \C$
     that are linearly independent over $\R$,
     we consider $\M$ as the quotient $\C/\mathbb{G}$. The underlying 
     identification is understood as the equivalence relation
     $z \sim z'$ between $z$ and $z'$ in $\C^n$ 
     whenever $z = z' +  l_1 \epsilon_1+ l_2 \epsilon_2$
     for some choice of $l \in {\mathbb{Z}}^{2}$. The resulting compact 
     manifold is  called a complex torus.
     Consider the space of holomorphic functions $\phi$ on $\C$
     that satisfy $\phi(z+\epsilon_1)=e^{-i \pi \omega_1} e^{\abs{\epsilon_1}^2/2}
     e^{\ol{\epsilon}_1 z} \phi(z)$ and 
     $\phi(z+\epsilon_2)= e^{-i \pi \omega_2} e^{\abs{\epsilon_2}^2/2}
     e^{\ol{\epsilon}_2 z} \phi(z)$, with $\omega_{1,2} \in \mathbb R$
     and $\frac{1}{\pi} {\Im}(\epsilon_1\ol\epsilon_2)=k \in \mathbb Z$
     which is seen to be a space of dimension $\abs k$ \cite{Bel61,Per86}.
     After picking a fundamental domain, one may identify  
     function values at equivalent points along the boundary 
     as coinciding vectors in a holomorphic fiber bundle. 
     The Hermitian metric from Example~\ref{ex:IIa}
     is compatible with this identification, so with $m$ the restriction
     of the Lebesgue measure to a fundamental domain 
     one arrives at a generalized Bergman space that
     does not result directly from the Lie-group setting considered by 
     Onofri \cite{Ono75}, but may be understood as an induced 
     representation, see \cite[Ch.~VIII]{Mau68} or \cite{Mac88},
     because this space would result from a change in the inner product of 
     Example~\ref{ex:IIa} and by demanding that the functions be invariant
     under the representation of a discrete abelian subgroup of the 
     Heisenberg-Weyl group.  
\end{enumerate}
\end{examples}

\subsection{Berezin-Toeplitz operators defined via quadratic forms}
\noindent
In the remaining text, we assume that the Hermitian metric $h$ 
and $\meas$ interpreted as a volume form are smooth and
non-degenerate to ensure that $\Bg$ is complete.
In addition, from now on all 
manifolds are tacitly assumed to be path-wise connected.

\begin{definition}
 Given the Hilbert space $\Bg$ and a real-valued function $f: \M \to \R$, 
we consider the sesquilinear form
\begin{align}
  \label{eq:Qform}
  \t_f: \Q( \t_f) \times \Q(\t_f)  &\longrightarrow \C \\
       (\psi, \phi) &\longmapsto \int_\M f(x) h_x(\psi(x), \phi(x)) d\meas(x) 
   \label{eq:tfdef}
\end{align}
with form domain 
\begin{equation}
  \label{eq:tfdom}
  \Q(\t_f) := \Bigl\{ \psi \in \Bg: \int_\M \abs{f(x)} h_x(\psi(x),\psi(x)) d\meas 
                                 < \infty \Bigr\} \, .
\end{equation}
When referring to $\t_f$ as a quadratic form, it is really the 
function $\psi \mapsto \t_f(\psi,\psi)$ that is meant.
\end{definition}

\begin{definition}
Whenever a real-valued function $f: \M \to \R$ gives rise 
to a semibounded closed form $\t_f\ge c, c \in \mathbb R$, it is  
associated with a unique self-adjoint operator $T_f$ satisfying 
$(\sqrt{T_f-c}\psi,\sqrt{T_f-c}\psi)=\t_f(\psi,\psi)- c (\psi,\psi)$
for all $\psi \in \Q(\t_f)$. In the context of generalized Bergman spaces, 
we call $T_f$ a self-adjoint Berezin-Toeplitz operator and 
the function $f$ its symbol.
\end{definition}

\begin{remarks} \label{rem:notdense} 
In analogy with the well-known KLMN theorem, see \cite{Sim71} or \cite[Thm.~X.17]{RS75},  
it is sufficient for the closedness 
and semiboundedness of $\t_f$ when the negative part
$f^-:=\max\{-f,0\}$ of $f$ can be incorporated as a perturbation 
of $\t_{f^+}$, $f^+:=\max\{f,0\}$, with relative form bound strictly less than one.
Even in this case it may be, due to singularities of $f$, that $\t_f$ 
is not densely defined and that in consequence, $T_f$ is self-adjoint only on the closure
of $\Q(\t_f)$ in $\Bg$.

The definition in terms of quadratic forms does not provide any
direct information about the domain of $T_f$ or how it operates on 
a given section. However, at least for bounded symbols $f$ we can give a more 
concrete description in which $T_f$ acts by its integral kernel.
\end{remarks}

\begin{definition}
A Schwartz kernel in a complex line bundle $\L$ is a family
of linear mappings $\{{\mathcal{S}}(x,y):$ $\L_y \to \L_x\}_{x,y \in \M}$, that
is, ${\mathcal{S}}(x,y)$ is linear in vectors with base point $y$ and has 
as its values vectors at $x$.   
If ${\mathcal{S}}(x,y)$ is jointly continuous in $x$ and $y$, then
it can be interpreted as continuous section  in the bundle 
$\L \otimes \L^* \to \M \otimes \M$,
where $\L^*$ is the dual bundle associating with each $x \in \M$
the space of complex linear forms on $\L_x$.
\end{definition}

\begin{remarks} \label{con:intkernels}
  If $h\meas$ is smooth and nowhere degenerate,
then the identity operator on $L^2_{\hol}(h\meas)$
a sesqui-analytic Schwartz kernel $K(x,y)$ \cite[Prop.~7]{Bod}.
Moreover, 
then any bounded operator $B$ on $L^2_{\hol}(h\meas)$ possesses a
sesqui-analytic Schwartz kernel
$B(x,y)$ that is characterized by the equation 
$h_x(u,B(x,y) v)= (K(\centerdot,x)u,$ $B K(\centerdot,y)v)$, $u \in \L_x$ and
$v \in \L_y$,
and the image of $\psi \in L^2_{\hol}(h\meas)$ is expressed as
\begin{equation}
  \label{eq:Bpsi}
  B \psi(x) = \int_\M   B(x,y) \psi(y) \, d\meas(y) \, .
\end{equation}

If $f$ is a bounded function, then the Schwartz kernel of the
operator $T_f$ is given by $h_x(u,T_f(x,y) v)= (K(\centerdot,x)u, f K(\centerdot,y)v)$
where the inner product is in $L^2(hm)$.

Since the right-hand side of equation \eq{Bpsi} 
is defined even for $\psi \in L^2(h\meas)$, any bounded 
operator extends naturally 
via its integral kernel to all of $L^2(h\meas)$. From this perspective,
$K(x,y)$ is the integral kernel of
the identity operator on $L^2_{\hol}(h\meas)$, which extends
to that of an orthogonal projection operator, 
henceforth called $K$, that maps $L^2(h\meas)$ onto $L^2_{\hol}(h\meas)$.
\end{remarks}

\subsection{Holomorphic maps between Riemann surfaces}
\noindent
In this subsection, we prepare the setting of the main result by
discussing how a holomorphic map onto a Riemann surface 
may be used to pull back the structures needed to define a 
generalized Bergman space over the domain of the map.

\begin{definition}
Given a Riemann surface $\M$,
any metric $g$ on $\M$ that is compatible with the almost complex structure
$J$ on $\M$ is called a conformal metric. By default, all metrics considered
are smooth.        
\end{definition}

\begin{itremark}
In a local coordinate system $z: U \to \C$, the compatibility requirement implies
that $g$ has the form
\begin{equation}
  \label{eq:localmetric}
  g = \frac{\gamma^2(z)}{2}  (dz\otimes d\ol z + d\ol z \otimes d z )
\end{equation}
with a conformal scaling function $\gamma: \C \to \{r > 0\}$.
The associated Dirichlet-Laplacian $\Delta$ is locally expressed as
\begin{equation}
  \label{eq:localDL}
  \Delta = \frac 4 {\gamma^2(z)} \frac{\partial^2}{\partial z \partial \ol z} \, ,
\end{equation}
where the abbreviations $\partial/\partial z := \half(\partial/\partial z_1 - i \partial/\partial z_2)$
and $\partial/\partial \ol z := \half(\partial/\partial z_1 + i \partial/\partial z_2)$
have been used. By inspection of \eq{localDL},
any linear combination of a holomorphic or antiholomorphic function is  harmonic
and vice versa.

As an aside, we remark that any smooth metric $g$  on an oriented surface
allows a complex analytic atlas for which $g$ is conformal \cite[Thm.~3.11.1]{Jos97}.
\end{itremark}

\begin{definition}
Let $\M$ and $\M'$ be Riemann surfaces with conformal metrics $g$ and $g'$, respectively.
A mapping $\Phi: \M' \to \M$ is called conformal if it is a local diffeomorphism
and $g_{\Phi(x')}(\Phi_* X',\Phi_* X') = \lambda^2(x') g'(X',X')$ 
holds for all $X' \in T_{x'}\M', x' \in \M'$ with a strictly positive 
dilatation function $\lambda: \M' \to \{r>0\}$.  
\end{definition}

\begin{lemma}
  Given two Riemann surfaces $\M$ and $\M'$ with conformal
metrics $g$ and $g'$ and a holomorphic map $\Phi$ 
from $\M'$ onto $\M$, then $\Phi$ is conformal
on the set where $\Phi_*$ is non-zero.
In addition, $\Phi$ is a harmonic morphism.
This means, a local harmonic function
$f: U \to \C$ defined on a chart domain $U \subset \M$ pulls back
to a harmonic function on $\Phi^{-1}(U)$.
\end{lemma}
\begin{proof} 
To prove this local property, it is enough to consider
the special case when both domains $U$ and $\Phi^{-1}(U)$
are open subsets of the complex number plane. 
The conformality of $\Phi$ results from that
of the metrics and because $\Phi$ satisfies the Cauchy-Riemann
differential equations.
By the chain rule
\begin{equation} \label{eq:loccd}
  \frac{\partial}{\partial z}\frac{\partial}{\partial \ol z} f \circ \Phi
 = \left( \frac{\partial}{\partial z}\frac{\partial}{\partial \ol z} f\right)
   (\Phi(z)) \frac{\partial \Phi}{\partial z}\frac{\partial \ol \Phi}{\partial \ol z}
\end{equation}
and the local form \eq{localDL} of the Dirichlet Laplacians associated with $g$ and $g'$,
$\Phi$ is a harmonic morphism.
\end{proof}

\begin{definition} \label{def:pb}
Let  $\M$ and $\M'$ be Riemann surfaces and suppose $\M$ is the base manifold
of a holomorphic line bundle $\L$. Given
a holomorphic map $\Phi$ from $\M'$ onto $\M$, then
the pull-back operation creates a bundle $\L'$ with fibers 
$\L_{x'}':=\pi^{-1}(\Phi(x')) \subset \dot {\bigcup}_{x' \in \M'} \L_{\Phi(x')}$ 
over $\M'$. 
The sections $\psi: \M \to \L$ transfer to $\L'$ 
by $\Phi^*\psi : x' \mapsto \psi (\Phi(x'))$, and 
the Hermitian structure $h$ on $\L$ pulls back to the fibers of $\L'$
by $\Phi^*h_{x'} := h_{\Phi(x')}$.

Let $\nabla$ denote the unique connection on $\L$
that is compatible with the complex and Hermitian structures \cite[Ch.~III]{Wel80}.
Then its pull-back, satisfying the identity
$(\Phi^* \nabla)_X \Phi^*\psi:=$ $\Phi^* (\nabla_{\Phi_* X} \psi)$
for smooth vector fields $X$ and sections $\psi$,
is in turn compatible with the complex and Hermitian structures 
present on $\L'$. In addition, the curvature form of $\Phi^* \nabla$ 
is the pull-back of the curvature on $\L$.
\end{definition}

\section{A Transformation Formula for Resolvents of
Be\-re\-zin-Toep\-litz Operators} \label{sec:3}

\subsection{Main result}

\begin{definition}
Let $\L$ be a holomorphic line bundle with a smooth, non-degenerate 
Hermitian metric $h$ on the fibers and a volume measure $m$ on the base
manifold. 
We denote the resolvent of a self-adjoint Berezin-Toeplitz operator
$T_f$ as $G^{hm}_{f-c}:=(T_f - c)^{-1}$
for any $c \in \C$ outside of the spectrum of $T_f$.
For such $c$ in the resolvent set, $G^{hm}_{f-c}$ is by definition a bounded operator
and via its integral kernel it extends according to Remarks~\ref{con:intkernels}
to all of $L^2(hm)$. 
In addition, if $T_f$ is not densely defined as mentioned in Remarks~\ref{rem:notdense},
we define $G^{hm}_{f-c}$ to be zero outside the 
closure of the domain $\cQ(\t_f)$ in $L^2_\hol(hm)$
of the sesquilinear form $\t_f$ corresponding to $T_f$.
In short, this extension is characterized by
$G^{hm}_{f-c} = G^{hm}_{f-c} K_f$ where
$K_f = K_f^* K_f$ is the orthogonal projector onto 
the closed subspace $\overline{\Q(\t_f)}$.
\end{definition}

\begin{definition}
Let $\M$ be a Riemannian complete manifold and 
$-\Delta\upharpoonright C_c^\infty(\M)$ the
self-adjoint negative Dirichlet-Laplace-Beltrami operator on $\M$.
A real-valued function $q: \M \to \R$ belongs to the
Kato class $\Kato$ if the following 
condition is satisfied:
\begin{equation}
  \label{eq:K_def}
  \lim_{t \downto 0} \sup_{x \in \M} \int_0^t \bigl(\ep{s \Delta} \abs{q}\bigr)(x)\, ds = 0 \; .
\end{equation}
Whenever this property holds only locally, which means for all products
$\chi_\Lambda q \in \Kato$ with characteristic functions $\chi_\Lambda$
of compact sets $\Lambda$ in $\M$, we write $q \in \Kloc$. 

If a real-valued function $q=q^+-q^-$, $q^\pm \ge 0$, satisfies $q^+ \in \Kloc$ and 
$q^- \in \Kato$ then it is called Kato decomposable, symbolized as $q \in \Katopm$.
\end{definition}

\begin{remarks}
  If the Ricci curvature of $\M$ is bounded below, then
bounds on the heat kernel imply the inclusion
$\Kloc \subset L^1_{\loc}(m)$, where $m$ is the Riemannian volume 
measure \cite{Dav85,Dav88,Dav89}. In this case, a 
 real-valued Kato-decomposable function $f: \M \to \R$ defines a 
semibounded Berezin-Toeplitz operator $T_f$ on $\Bg$ 
that is obtained from the closure $T_f\upharpoonright K_f(C_{c\L}^\infty(\M))$
where $C_{c\L}^\infty(\M)$ denotes the space of smooth, compactly
supported sections \cite[Thm.~29]{Bod}. 
\end{remarks}

\begin{theorem} \label{thm:main}
Let $\M$ and $\M'$ be two Riemannian complete surfaces equipped with 
conformal metrics $g$ and $g'$,
and suppose $\M$ is the base manifold of a holomorphic line bundle $\L$.
Furthermore, let $\Phi: \M' \to \M$ be a holomorphic, surjective 
mapping with dilatation function $\lambda$, that is,
$g'=\lambda^2 \Phi^*g$. The pull-back bundle $\L':=\Phi^*\L$
is thought of as being equipped with the Hermitian metric $h':=\Phi^*h$.
Assume that the Ricci curvatures of $\M$ and $\M'$ are bounded below and that 
the functions
$f: \M \to \R$ and $\lambda^2 f \circ \Phi: \M' \to \R$ are
Kato decomposable such that the corresponding Berezin-Toeplitz operators can be
defined via semibounded quadratic forms on $L^2_\hol(hm)$ and $L^2_\hol(h'm')$,
where $m'$ is the natural volume with respect to $g'$. 

If $\psi \in L^2_\hol(hm)$ and $\lambda^2 \psi\circ\Phi \in L^2(h'm')$,
then there is a relationship between the resolvents
\begin{equation}
  \label{eq:Rreln}
 \left(G^{hm}_{f-c} \psi\right)(\Phi(x')) =
  \left(G^{h'm'}_{\lambda^2 (f \circ \Phi - c)} \lambda^2 \psi \circ \Phi\right)(x')
  \, 
\end{equation}
for such $c \in \C$ that satisfy the operator inequalities $T_f > {\Re} c$ and 
$T_{\lambda^2 f \circ \Phi} > {\Re} c \, T_{\lambda^2}$. 
Hereby, the extension convention according to Remarks~\ref{con:intkernels} 
is implicit, since $\lambda^2 \psi\circ\Phi$
is not holomorphic unless $\lambda$ is constant.
If additionally $\psi\circ\Phi \in L^2_\hol(h'm')$, then
this equation reads
\begin{equation}
  \label{eq:RrelnII}
 \left(G^{hm}_{f-c} \psi\right)(\Phi(x')) =
  \left(G^{h'm'}_{\lambda^2 (f \circ \Phi - c)} T_{\lambda^2} \psi \circ \Phi\right)(x')
  \, .
\end{equation}
\end{theorem}

\begin{remarks} \label{rem:maybeconst}
It is a nontrivial issue to establish that the values for $c$
allowed by the two operator inequalities in the assumptions of the preceding theorem
form a nonempty open set in $\C$. The first inequality is easy to satisfy
for sufficiently negative values of ${\Re} c$. However, the only generally sufficient
condition known to the author to ensure the second inequality is
$T_{\lambda^2} \ge \varepsilon >0$, with again a sufficiently negative
$\Re c$. This can be deduced, for example, if $\M$ is compact, since the singular
points of $\Phi$ are isolated \cite{Fug78} and thus $T_{\lambda^2} \ge \inf\{\int_\M \rho \lambda^2 dm:$ 
$\rho \ge 0, \int_\M \rho dm =1\}>0$.   

  It may happen that $c'=\lambda^2 f \circ \Phi$ is constant, see
the following examples.
In this case, formula~\eq{RrelnII} simplifies to
\begin{equation}
  \label{eq:RrelnIII}
 \left(G^{hm}_{f-c} \psi\right)(\Phi(x')) =
  \left(G^{h'm'}_{c'-c\lambda^2} \lambda^2 \psi \circ \Phi\right)(x')
  \, .
\end{equation}
and if $T_{\lambda^2}$ is known, the left-hand side can be computed explicitly .
It is interesting to note that in a sense, $f$ and $\lambda^2$ switch roles:
The constant in the resolvent becomes a multiplier of the emerging 
symbol $\lambda^2$ and the former symbol $f$ gets turned into a constant.
In this special case,
the validity of the second operator inequality in the assumptions 
can then be established in yet another way: If $0 \neq f \in \Kato$ then
$T_f$ is bounded and one may flip the sign of $f$ to ensure that
the resulting constant $c'$ is strictly positive, and as a consequence equation \eq{RrelnII}
holds for any $c$ from the open half-plane $\{c: {\Re} c<T_f\}$.

Once an open set for the allowed values of $c$ in equation~\eq{RrelnIII}
is established,
one can extend this set by analytic continuation of the resolvents. We follow the 
nice exposition by Wittich \cite{Wit00} of the facts
collected from Kato's book \cite{Kat76}.
Hereby, it is important that for $u \in \Katopm$, 
$\{T_{\xi u}\}$ forms a holomorphic
family of type~B with $\xi \in {\C}, {\Re} \xi >0$ \cite[Sec.~VII.2]{Kat76} 
and that for $\xi$ in 
a smaller, compact set $C$ in the right half-plane with nonempty interior, the
numerical ranges of all $T_{\xi u}$ are 
contained in a
common sector of complex numbers with real parts bounded below 
by a constant $\zeta<0$ \cite[Thm.~VII.4.2]{Kat76}. 
In addition, the set $\{(\xi,c) \in {\C}^2: \xi \in C, {\Re} c <\zeta\}$
is a set of holomorphy for $(\xi u - c)^{-1}$ \cite[Ch.~V]{Kat76}. Since
the resolvents of holomorphic families of type~B have unique 
analytic continuations, one may then extend the set of values of $c$ 
to all values for which both resolvents in equation~\eq{RrelnIII}
exist \cite[Rem.~VII.1.6]{Kat76}.
 
At first, the generalization of the theorem to higher dimensions of $\M$ and $\M'$ seems
straightforward. Unfortunately, one does not gain more generality by restating it
this way, because if $\dim_\C \M' = \dim_\C \M \ge 2$, then the
harmonic morphism
$\Phi$ is necessarily a local isometry up to an overall rescaling by a constant
\cite{Fug78}. Thus, the result in higher dimensions concerns solely covering 
maps of K\"ahler manifolds. 
The case $\dim_\C \M' > \dim_\C \M$ is not interesting in
the quantization context because
the pull back of the curvature would be degenerate; in other words,
the correspondence principle would lead to a classical system with
a degenerate symplectic form.

Finally, we remark that
the statement of the theorem does not refer directly to the probabilistic
elements used in the proof given hereafter. It would be nice to find an
alternative derivation of the claimed relationship with a purely
analytic argument.
\end{remarks}

\begin{examples}
The following examples have the virtue that, at least in special
cases, one may verify the formulas by other means than the probabilistic
derivation.
  \begin{enumerate}
  \item A whole class of examples is given whenever $(\M',\Phi)$ defines a covering 
   of $\M$. Then there is a unique complex structure on $\M'$ such that
   $\Phi: \M' \to \M$ is holomorphic \cite[Prop.~5.8.3]{Gol98}, and
   the conformal metric $g$, the bundle $\L$, and the 
   Hermitian metric $h$ 
   may be pulled back to give a complex manifold $\M'$ equipped with $g'$ 
   that is the base manifold of the Hermitian holomorphic line bundle 
   $\L'=\Phi^*\L$.
   The dilatation is just $\lambda=1$ because the map $\Phi$ is
   a local isometry. Thus, the formula
   \begin{align}
       (G^{hm}_{f-c}\psi)(\Phi(x')) = (G^{h'm'}_{f\circ \Phi-c} \psi\circ\Phi)(x')
   \end{align}
   relates the resolvent of $T_f$ to that of its periodic extension $T_{f \circ \Phi}$ 
   on the covering space $\M'$. For a more concrete example 
   of this situation and the details of a relation 
   between operators on the spaces in Examples~\ref{ex:I}.\ref{ex:IIa} and
   \ref{ex:I}.\ref{ex:theta}, see \cite{BK01}.
  \item Let $\M=\M'=\C$ be equipped with the standard metric
   $g=g'=\frac 1 2 (dz \otimes d\ol z + d\ol z \otimes dz)$
   and take $\L$ and $\L'$
   to be the trivial bundles $\C \times \C$. Suppose $\L$ is
   equipped with the Hermitian metric from the Barut-Girardello space 
   described in Examples~\ref{ex:I}.\ref{ex:II}.
   The mapping $\Phi: z \mapsto z^2$ from $\M'$ to $\M$ then
   pulls back the Hermitian metric so that $\L'$ becomes the
   bundle underlying the
   Fock-Bargmann space, Examples~\ref{ex:I}.\ref{ex:IIa}.
   The square of the dilatation is given by $\lambda^2(z)=4 \abs{z}^2$.
   Just as mentioned in Remarks~\ref{rem:maybeconst}, choosing
   $f(z)=c'/4\abs{z}$ with a constant $c'\in \mathbb C$ then leads to 
   $\lambda^2(z) f(\phi(z)) =  c'$. In short, the resolvent
   relation
   \begin{align}
     \Bigl(G^{hm}_{\frac{c'}{4\abs z}-c} \psi\Bigr)(\Phi(z))
   = \Bigl(G^{h'm'}_{c'-c\abs{z}^2} T_{\abs{z}^2} \psi\circ\Phi\Bigr)(z)
   \end{align}
   is derived, a priori valid for $c'$ with a sufficiently negative 
   real part and $c$ with a sufficiently large positive real part.
   Indeed, one may verify that the spectra of the Berezin-Toeplitz
   operators that are related here are given by
   $\mathop{\mathrm{spec}}(T_{\frac{c'}{4\abs z}})=\{\frac{c'}{8n+4}: n \in {\mathbb N}\}\cup\{0\}$
   and $\mathop{\mathrm{spec}}(T_{c \abs{z}^2})=c \mathbb N$, due to their
   known eigensections that are just the monomials $z \mapsto (z,z^n)$, $n \in \mathbb N$.
   Of course, the eigensection decomposition offers another method to verify the resolvent 
   relation.

   The mapping $\Phi: z \mapsto z^2$ is of the Clifford type \cite{Bai90} and represents 
   the direct analog on two-dimensional Euclidean space of the harmonic morphism discussed 
   in the example of \cite{Wit00}. There, the resolvent of the  Coulomb system in 
   dimension 3 is related to that of the harmonic oscillator in dimension 4.    
   In a similar vein, here the components $\Phi_1(z)=z_1^2 - z_2^2$
   and $\Phi_2(z)= 2 z_1 z_2$ of the mapping $\Phi$ may be 
   interpreted as quadratic forms on ${\mathbb R}^2$, and the symmetric matrices corresponding 
   to these quadratic forms are the basis of a two-dimensional Clifford
   algebra over~$\R$. According to the common scheme behind both examples, the anticommutation 
   relations observed by the matrices imply that $\Phi$ is harmonic and horizontally conformal 
   at all non-singular points. Since $\Phi$ is also surjective, it is a harmonic 
   morphism \cite{Fug78}. 
\item Choose $\M = \M'=\C \mathrm{P}^1$. We will focus on one chart
   which maps all but one point of each manifold stereographically
   to the complex plane. In these local coordinates let the Riemannian conformal
   metrics be given as $g=g'=\half \frac{1}{(1+\abs{z}^2)^2} (dz \otimes d\ol z +
   d\ol z \otimes dz)$. Let $\M$ be the base manifold of the bundle 
   described in Examples~\ref{ex:I}.\ref{ex:IIb}, so in a local
   trivialization the holomorphic, square-integrable sections in $\Bg$ 
   are given by polynomials of maximal degree $k\in {\mathbb Z}^+$, where the
   Hermitian metric is $h_z((z,u),(z,v))=\frac{1}{(1+\abs{z}^2)^k} \ol u v$.
   Take the conformal map $z \mapsto \alpha z + \beta$ with $\alpha, \beta \in \C$
   and $\alpha \neq 0$ that fixes the point excluded from the chart domain.
   Then the square of the dilatation is $\lambda^2(z)=\abs{\alpha}^2 
   \frac{(1+\abs{z}^2)^2}{(1+\abs{\alpha z + \beta}^2)^2}$. The 
   pull-back of the Hermitian metric 
   $h'_z((z,u),(z,v)=\frac{1}{(1+\abs{\alpha z + \beta}^2)^k} \ol u v$
   together with the Riemannian volume $m'$ define a new Hilbert
   space $L^2_\hol(h'm')$. Taking $f(z)= \frac{c'}{\abs{\alpha}^2} 
   \frac{(1+\abs{z}^2)^2}{(1+\abs{z-\beta}^2/\abs{\alpha}^2)^2}$
   yields the resolvent relationship
   \begin{align} \phantom{\,}
     \Bigl(G^{hm}_{ \frac{c'}{\abs{\alpha}^2} 
   \frac{(1+\abs{z}^2)^2}{(1+\abs{z-\beta}^2/\abs{\alpha}^2)^2}-c} \psi\Bigr)(\Phi(z)) =
     \Bigl(G^{h'm'}_{c'- c \abs{\alpha}^2 
   \frac{(1+\abs{z}^2)^2}{(1+\abs{\alpha z + \beta}^2)^2}} \psi\circ\Phi\Bigr)(z) \, ,
   \end{align}
   valid again for  $c'$ with a sufficiently negative 
   real part and $c$ with a sufficiently large positive real part.
   In case $\beta = 0$  the eigensections are again given by monomials and
   one may verify that the eigenvalues of the corresponding 
   Berezin-Toeplitz operators are inverses to each other,
   $\mathop{\mathrm{spec}}(T_{f})= \{ \abs{\alpha}^{2n} \/_2F_1(k$, $n+1,2+k;1-\alpha):
   n = 0, 1, \dots k\}$ and
   $\mathop{\mathrm{spec}}(T_{\lambda^2})= \{ (\abs{\alpha}^{2n} 
    \phantom{}_2F_1(k$, $n+1,2+k;1-\alpha))^{-1}: n = 0, 1, \dots k\}$.
   Indeed, this inverse relationship is to be expected also for $\beta\neq 0$ 
   by observing that the conformal map may simply be interpreted as a coordinate 
   transformation that turns the sesquilinear form of one operator into the 
   inner product on the other space.
  \end{enumerate}
\end{examples}

\subsection{Assembly of the proof}
\noindent
The major ingredients of the proof of Theorem~\ref{thm:main}
are a representation of resolvents of Berezin-Toeplitz
operators in the form of so-called Wiener-regularized path integrals
and an invariance property of Brownian motion under harmonic
morphisms that implies a simple substitution rule
in the path-integral representation. Before the final assembly,
we explain the ingredients.

\subsubsection{Probabilistic representation of resolvents 
of Berezin-Toeplitz operators}

\begin{definition}
  We adopt the usual terminology: 
An almost surely continuous process $\sB$ with values in 
the Riemannian manifold $\M$ is called Brownian
motion with diffusion constant $D>0$
if for every smooth function $\phi \in \Cinfty(\M)$, the difference
\begin{equation}
  \label{eq:BM}
  \sM_t := \phi\circ \sB_t - \phi \circ \sB_0 -  \int_0^t D \Delta \phi \circ \sB_s ds
\end{equation}
is a real-valued continuous local martingale $\sM$.

A probability measure governing Brownian motion with a 
diffusion constant $D>0$ and almost
surely fixed starting point $\sB_0=x$
will be denoted as $\bP^D_x$. The expectation
with respect to this probability measure
is written as $\E^D_x$.
\end{definition}

\begin{itremark}
    A complete Riemannian manifold $\M$ 
with Ricci curvature bounded below is
Brownian complete, that is, for
a fixed diffusion constant $D>0$, 
a Brownian motion $\sB$ starting at any $x \in \M$
has an infinite explosion time \cite[Ch.~V]{Eme89}. 
\end{itremark}

\begin{proposition}
Let $\L$ be a Hermitian holomorphic line bundle over a Riemann surface $\M$
that is Riemannian complete with Ricci curvature bounded below. Denote
by $m$ the natural volume measure on $\M$ and by $\bP^D$ a family of
Brownian-motion measures having a common diffusion constant $D>0$.
Let the real-valued function $\rho$ on $\M$ be specified by the  
curvature term $\rho(x) \psi(x) = (\nabla_{\ol Z} \nabla_{Z} - \nabla_{Z}  \nabla_{\ol Z}
 - \nabla_{\ol Z Z - Z \ol Z})\psi(x)$ 
with any smooth, locally non-vanishing section $\psi$ and
an arbitrary choice of a holomorphic tangent vector field $Z$ 
that is normalized at $x$, that is, $g_x(\ol Z,Z)=1$  
in terms of the bilinear extension of $g$ to the complexified tangent space.

If $f,\rho \in \Katopm$ then the image of a section $\psi \in \Bg$
under the semigroup $e^{-t T_f}K_f$ is for $t>0$ given by
the ultra-diffusive limit of a Brownian-motion expectation,
\begin{equation}
  \label{eq:DKformula}
  \bigl(e^{-tT_f}K_f\bigr)\psi (x) = \lim_{D \to \infty} 
    \E_{x}^{D}\Bigl[ \ep{-\int_0^t (D\rho(\sB_r) + f(\sB_r))dr}
                                                      H^{-1}_{\sB,t} \psi(\sB_t)\Bigr] \, .
\end{equation}
The inverse $H^{-1}_{\sB,t}$ of the stochastic horizontal transport appearing in this 
so-called Wiener-regularized path integral is associated with the compatible connection $\nabla$ 
and therefore preserves the length of a transported vector. The reverse transport
can either be understood by appealing to localized expressions \cite{Sch80},
i.e.\ one restricts to a subspace of the probability space 
by introducing exit times of local coordinate patches and then reformulates
the reverse horizontal transport in a local trivialization, or one interprets 
\eq{DKformula} as a shorthand for
\begin{equation}
  \label{eq:heatLeqnII}
  h_x(u, e^{-tT_f} K_f \psi )
 = \lim_{D \to \infty} 
   \E^D_x\left[e^{-\int_0^t (D\rho(\sB_r) +f(\sB_r)) dr} h_{\sB_t}(H_{\sB,t} u,\psi(\sB_t))\right] \,  
\end{equation}
with an arbitrary reference vector $u \in \L_x$.
\end{proposition}
\begin{proof} 
The detailed proof is given in \cite[Thm.~45]{Bod}.
We summarize the key ingredients.

To begin with, we consider a version of the Feynman-Kac formula for 
perturbations of the semigroup generated by the Bochner Laplacian 
\cite[Ch.~IX]{Bis81} and the limiting argument in 
\cite[App.~C]{Bod} that permits a larger class of perturbations. 
In combination with a Weitzenboeck-type formula one then concludes 
that the pre-limit expression on the right-hand
side of equation \eq{DKformula} is the image of $\psi$ under
the semigroup generated by the Schr\"odinger operator 
$S_{D,f}^{(0,\bullet)} := - D \Delta^{(0,\bullet)} + f$.
This operator is understood as the form sum $- D \Delta^{(0,\bullet)} + f^+$
that is perturbed by the negative part $f^-$. The non-negative
operator $-  \Delta^{(0,\bullet)}$, in turn, is the form closure
of a differential operator initially defined by
 $-\Delta^{(0,\bullet)} \psi(x) = - \nabla_Z \nabla_{\ol Z} \psi(x)
- \nabla_{\Cov_Z \ol Z} \psi(x)$ on smooth, compactly supported sections $\psi$,
whereby $\Cov$ denotes the Levi-Civita connection and
$Z$ is again an arbitrary locally non-vanishing, holomorphic
vector field that is normalized at~$x$.

Now the proof is seen to be equivalent to showing the pointwise identity
\begin{equation}
   \label{eq:sgconv}
   \bigl(\ep{- t T_f} K_f \psi \bigr)(x)
   = \lim_{D \to \infty} \ep{-t  S^{(0,\bullet)}_{D,f}} \psi (x)\, .
\end{equation}
This is implied by continuity properties of both sides and
strong convergence of $\ep{-tS^{(0,\bullet)}_{D,f}}$ 
to $\ep{- t T_f} K_f$
in the limit $D \to \infty$, a
result of monotone convergence of the forms associated
with $\{S_{D,f}^{(0,\bullet)}\}_{D>0}$ and the fact that
$-  \Delta^{(0,\bullet)}$ vanishes only on holomorphic sections
and is otherwise strictly positive.
\end{proof}

\begin{lemma}  \label{prop:Grepp}
  If for some $D>1$, ${\Re} c < S^{(0,\bullet)}_{D,f}$, then 
in analogy with \eq{sgconv} we obtain the integral representation
for the resolvent
\begin{equation}
  \label{eq:Gintrepnpwise}
  G^{hm}_{f-c} \psi (x)
  = \lim_{D \to \infty} 
     \int_0^\infty \ep{-t S^{(0,\bullet)}_{D,f} + t c} \psi(x) \, dt \, .
\end{equation}
\end{lemma}
\begin{proof} 
The strong convergence
\begin{equation} \label{eq:19}
   G^{hm}_{f-c} \psi 
  = \lim_{D \to \infty} 
     \int_0^\infty \ep{-t S^{(0,\bullet)}_{D,f} + t c} \psi \, dt \, 
\end{equation}
 is again a result of the monotone convergence of forms
associated with $S^{(0,\bullet)}_{D,f}$.
To obtain the pointwise equality, we apply the point-evaluation functional 
 $\vartheta_u=h_x(u,\centerdot)$ to the integrand of \eq{Gintrepnpwise},
use the self-adjointness and semigroup property to obtain
\begin{equation}
     h_x(u,\ep{-t S^{(0,\bullet)}_{D,f}} \psi (x))
   = ( \ep{-\frac t D S^{(0,\bullet)}_{D,f}}(\centerdot,x)u,
                   \ep{-t (1-\frac 1 D) S^{(0,\bullet)}_{D,f}} \psi) \, ,
\end{equation}
and integrate over $t \in [0,\infty)$, which yields
a pointwise expression for the image of $\psi$ under the resolvent
of $S^{(0,\bullet)}_{D,f}$:
\begin{equation} \label{eq:21}
  h_x(u, (S^{(0,\bullet)}_{D,f} - c)^{-1} \psi(x)) 
 = \int_0^\infty ( \ep{-\frac t D S^{(0,\bullet)}_{D,f}}(\centerdot,x)u,
                   \ep{-t (1-\frac 1 D) S^{(0,\bullet)}_{D,f} + tc} \psi) \, dt \, .
\end{equation}
The uniform boundedness of 
$\ep{-t (1-\frac 1 D) S^{(0,\bullet)}_{D,f}}$ in $D$ and the strong convergence
of the function $\ep{-\frac t D S^{(0,\bullet)}_{D,f}}(\centerdot,x)u$ to 
$\ep{t \Delta^{(0,\bullet)}}(\centerdot,x)u$ \cite[Lem.~44]{Bod}
imply that in the limit $D \to \infty$ the integral on the right-hand side of equation~\eq{21} converges to
\begin{equation}
  h_x(u, G^{hm}_{f-c} \psi(x) ) =
  h_x(\ep{t \Delta^{(0,\bullet)}}(\centerdot,x)u, G^{hm}_{f-c} \psi) \, 
\end{equation}
which proves the claimed identity.
\end{proof}

\subsubsection{An invariance property of Brownian motion}

\begin{definition} Let $\sB$ be a Brownian motion on a Riemann surface $\M$.
An additive functional of Brownian motion is a stochastic process $\sA$ given 
in the form
\begin{equation}
  \label{eq:addfnal}
  \sA_\sigma := \int_0^\sigma q(\sB_s) ds \, 
\end{equation} 
with a non-negative function $q: \M \to \R^+$.

If $\sA$ is everywhere finite, increasing without jumps, and if
$\lim_{\sigma \to \infty} \sA_\sigma = \infty$ with probability one,
then we define a stochastic time change by the inverse $\tau$
of $\sA$, in other words, $\sA_{\tau(t)}=t$ for all $t \ge 0$.
\end{definition}

\begin{lemma}
Let $\M$ and $\M'$ be two Riemann surfaces with conformal metrics
$g$ and $g'$. Suppose both manifolds are Riemannian complete
with Ricci curvatures bounded below.
Given $\sB'$ a Brownian motion on $\M'$ and
$\Phi: \M' \to \M$ a surjective holomorphic mapping
having the dilatation function $\lambda: \M' \to \R^+$,
then the additive functional
$\sA_\sigma := \int_0^\sigma \lambda^2(\sB'_s) ds$
satisfies the finiteness and limit conditions in the preceding definition
of the stochastic time change $\tau$ and
$\{\sB_t:=\Phi(\sB'_{\tau(t)})\}_{t \ge 0}$
defines a Brownian motion $\sB$ on $\M$. 
\end{lemma}
\begin{proof} 
Essential to the proof is that since $\Phi$ is holomorphic and surjective, its singular points
are isolated and thus by an argument of Fuglede \cite{Fug78} polar.
Therefore, $\lambda(\sB')$ is an a.s.\ strictly positive, continuous process
and $\sA$ is increasing. 
Implicit in the assumptions is that $\M'$ is Brownian complete, so  
$\{\Phi(\sB'_{\tau(t)})\}_{t \ge 0}$ 
is some stochastic process with values in $\M$ that may possibly have finite
values for its explosion time.
However, since $\Phi$ is a harmonic morphism, it relates the generators of 
Brownian motion by a conformal scaling operation \cite{CO83}
as in the localized version \eq{loccd}. Therefore, the time-changed
process is by the characterization \eq{BM}
seen to be a Brownian motion on $\M$ so that due to the lower
bound on the Ricci curvature it has an infinite
explosion time.
\end{proof}

\subsubsection{Implementing the substitution rule in the probabilistic representation}

\begin{proof}[Proof of Theorem~\protect{\ref{thm:main}}]
Since we can always absorb the constant $z$ into the definition of $f$,
we will for convenience of notation assume $z=0$.
  Using the probabilistic representation, we have
  \begin{equation}
    \label{eq:PResrep}
    \bigl(G^{hm}_f \psi\bigr)(\Phi(x')) 
    = \lim_{D\to \infty} \int_0^\infty 
                              \E^D_{\Phi(x')}\Bigl[\ep{-\int_0^t (D\rho + f)(\sB_s)ds}
                               H^{-1}_{\sB,t}\psi(\sB_t)\Bigr] 
                         dt  \, 
  \end{equation}
whenever $f$ leads to a Berezin-Toeplitz operator 
the real part of which is bounded below by zero.
By the invariance property of Brownian motion, we can replace 
$\sB_s$ with $\Phi(\sB_{\tau(s)})$,
\begin{multline}
  \label{eq:chgvar}
  \bigl(G^{hm}_f \psi\bigr)(\Phi(x')) 
  = \lim_{D\to \infty} \int_0^\infty \E^D_{x'}\Bigl[\ep{-\int_0^t (D\rho\circ\Phi
                                                                 + f\circ\Phi)(\sB_{\tau(s)}) ds}\\
                              H^{-1}_{\Phi(\sB),\tau(t)}\psi\circ\Phi(\sB_{\tau(t)})\Bigr] dt  \, .
\end{multline}
Interchanging the integration with the expectation and substituting
gives
\begin{multline}
  \label{eq:subsresult}
  \bigl(G^{hm}_f \psi\bigr)(\Phi(x')) 
 =  \lim_{D\to \infty}
    \E^D_{x'}\Bigl[ \int_0^\infty \!\!\!\ep{-\int_0^\tau\!  
     \lambda^2 (\sB_\sigma) (D \rho\mkern-2.7mu\circ\mkern-3mu\Phi 
      \mkern-1mu+\mkern-1mu f\mkern-3mu\circ\mkern-3mu\Phi)(\sB_\sigma) d\sigma} \\
       \lambda^2(\sB_\tau) \Phi^* \mkern-1mu H^{-1}_{\sB,\tau}\psi\circ\Phi(\sB_{\tau}) d\tau \Bigr] \, ,
\end{multline}
where we have used that, as a consequence of the local properties under the pull-back
operation explained in Definition~\ref{def:pb},
the horizontal transport and pull-back
operations commute in the following sense: Suppose we lift a 
(Brownian) path in $\M$ to $\M'$. Then first horizontally transporting a 
vector $u$ through the fibers along the path in $\M$ and lifting it to $\L'$
gives the same as lifting it first and then horizontally transporting
along the lifted path in $\M'$.

After reversing the order of integration again, one may interpret the resulting
expression as the integral representation of a new resolvent, in short
\begin{equation}
  \label{eq:ffinal}
  \bigl(G^{hm}_f \psi\bigr)(\Phi(x')) = \bigl(G^{h'm'}_{\lambda^2 f\circ \Phi} \, \lambda^2 \psi\circ \Phi \bigr)(x') \, ,
\end{equation}
because of the identity $\rho'(x') = \lambda^2(x') \rho(\Phi(x'))$
between the curvature terms that results from the conformal relationship 
between $g'$ and $g$ and the definition of the pull back of the bundle curvature.
However, to ensure the validity of this integral representation one depends on the 
assumption of the second operator inequality in the statement of Theorem~\ref{thm:main}.
\end{proof}

\paragraph{{\bf Acknowledgements}}
Thanks are extended to Michael Aizenman and John Klau\-der for helpful remarks
and to H\'el\`ene Rey for sharing her enthusiasm about mathematical physics, an 
un\-paralleled source of encouragement and inspiration.

\end{document}